\documentstyle[12pt]{article}
\begin{document}

\noindent Stockholm\\
2007\\

\vspace{15mm}

\begin{center}

{\Large NOTE ON NON-METRIC GRAVITY}

\vspace{15mm}

{\large Ingemar Bengtsson}\footnote{Email address: ingemar@physto.se. 
Supported by VR.}

\vspace{5mm}

{\sl Stockholm University, AlbaNova\\
Fysikum\\
S-106 91 Stockholm, Sweden}

\vspace{15mm}

{\bf Abstract}

\end{center}

\vspace{6mm}

\noindent We discuss a class of alternative gravity theories that are specific 
to four dimensions, do not introduce new degrees of freedom, and 
come with a physical motivation. In particular we sketch their Hamiltonian 
formulation, and their relation with some earlier constructions.

\newpage

\noindent A class of alternative gravity theories were recently introduced, 
under the name of ``non-metric gravity'' \cite{Kirill}. There are reasons to take this 
class seriously, in particular arguments were advanced why this class---which 
is defined by one free function of two variables---should be closed under 
renormalization \cite{Krasnov}. Indeed the construction is interesting already on the 
classical level, since it does not introduce any new degrees of freedom, 
as compared to GR (which is a member of the class); moreover the construction 
is intrinsically four dimensional (arguably a good thing), and is based 
on a clean split between conformal structure and conformal factor of the metric. 
In fact the split is so clean that the latter is largely lost track of, which 
brings us to some weak points of these models. The first of those is that the 
models describe complex spacetimes, and it is not clear how to recover the 
Lorentzian sector. The second is that they appear to be quite difficult to 
couple to matter (but this may turn out to be a strong point in the 
end). 

The purpose of this note is to clarify the relation between non-metric gravity 
and an earlier class of models with similar properties \cite{Peter,Ingemar}. 
The starting point is Pleba\'nski's action for vacuum general relativity (GR) 
\cite{Plebanski,Mason,Reisenberger}:

\begin{equation} S = \frac{1}{4} \int d^4x \ \epsilon^{\alpha \beta \gamma \delta}
\left[ \Sigma_{\alpha \beta i}F_{\gamma \delta i} - \frac{1}{2}\Phi_{ij}
\Sigma_{\alpha \beta i}\Sigma_{\gamma \delta j}\right] \ , \label{krasnov} 
\end{equation}

\noindent where $\Phi$ is a symmetric traceless three by three matrix, 

\begin{equation} F_{\alpha \beta i} = \partial_\alpha A_{\beta i} - \partial_\beta 
A_{\alpha i} + if_{ijk}A_{\alpha j}A_{\beta k} \ , \end{equation}

\noindent and $f_{ijk}$ are the structure constants of the SU(2) Lie algebra. 
To discuss the corresponding space of solutions, introduce a spacetime 
metric given in terms of a tetrad of vector fields by 

\begin{equation} g_{\alpha \beta} = \eta_{IJ}e_{\alpha}^Ie_{\beta}^J \ , 
\end{equation}

\noindent where $\eta_{IJ}$ is the Minkowski metric. Use the tetrad to 
form a 2-form 

\begin{equation} \sigma_{\alpha \beta}^{IJ} = e_{[\alpha}^Ie_{\beta ]}^J \ . 
\end{equation}

\noindent One can show that, in any solution of the equations coming from 
Pleba\'nski's action, the 2-form $\Sigma_{\alpha \beta i}$ is 
the self dual part of a 2-form $\sigma_{\alpha \beta}^{IJ}$, 
the metric $g_{\alpha \beta}$ solves Einstein's vacuum equations, the connection 
$A_{\alpha i}$ is the metric compatible self dual spin connection, and 
$\Phi_{ij}$ is the self dual part of the Weyl tensor---which is indeed 
naturally a symmetric traceless complex three by three matrix \cite{Ehlers}, 
even if it is more common to regard it as a rank 4 spinor. We will 
not explain these matters further here---the argument given below will lead to 
the same conclusion in a different way.

Krasnov modifies this action by setting 

\begin{equation} \Phi_{ij} = \phi_{ij} + \delta_{ij}\Lambda(\mbox{Tr}\phi^2, 
\mbox{Tr}\phi^3) \ , \label{kirill} \end{equation}

\noindent where $\mbox{Tr}\phi = 0$. If the function $\Lambda$ is constant this 
results in Einstein's theory including a cosmological constant. Otherwise we have 
a new class of models, not equivalent to GR. 
 
Although we are motivated by our belief that Krasnov's construction is connected 
to some deep mathematical properties of four dimensional manifolds, the only 
mathematics we need for now is the characteristic equation for three by 
three matrices, namely 

\begin{equation} M^3 - M^2\mbox{Tr}M - \frac{1}{2}M\left(\mbox{Tr}M^2 - 
(\mbox{Tr}M)^2\right) - \det{M} = 0 \ . \end{equation}

\noindent It follows that 

\begin{equation} \mbox{Tr}M^3 - \frac{3}{2}\mbox{Tr}M^2\mbox{Tr}M + 
\frac{1}{2}(\mbox{Tr}M)^3 - 3\det{M} = 0 \ . \end{equation}

\noindent A three by three matrix has three independent invariants only. Thus 
equipped, we can proceed to cast Krasnov's models into Hamiltonian form. 
Afterwards we will make some further comments, in particular we will discuss 
how a spacetime metric enters the game.

We begin. Vary the action (\ref{krasnov}) with respect to the 2-form 
$\Sigma_{\alpha \beta i}$, solve for the same, and insert the result back 
into the action. The result is 

\begin{equation} S = \frac{1}{8}\int d^4x \ \epsilon^{\alpha \beta \gamma 
\delta}\Psi_{ij}F_{\alpha \beta i}F_{\gamma \delta j} \ ,  
\label{lee} \end{equation} 

\noindent where 

\begin{equation} \Psi \equiv \Phi^{-1} = \frac{3(\phi^2 - \phi\Lambda - 
\frac{1}{2}\mbox{Tr}\phi^2 + \Lambda^2)}{\mbox{Tr}\phi^3 - 
\frac{3}{2}\mbox{Tr}\phi^2\Lambda + 3 \Lambda^3} \ . \end{equation}

\noindent Here we used the characteristic equation, and we closed our eyes to 
the fact that $\Phi$ may be non-invertible---this happens when its algebraic 
Petrov type is $\{ - \}$, $\{ 4\}$, or $\{ 3,1\}$. The invariants of 
the matrix $\Psi$ are given by 

\begin{equation} I_1 \equiv (\mbox{Tr}\Psi )^2 - \mbox{Tr}\Psi^2 = \frac{18\Lambda}
{\mbox{Tr}\phi^3 - \frac{3}{2}\mbox{Tr}\phi^2\Lambda + 3\Lambda^3} 
\label{i1} \end{equation}
 
\begin{equation} I_2 \equiv \mbox{Tr}\Psi = 
\frac{3(3\Lambda^2 - \frac{1}{2}\mbox{Tr}\phi^2)}
{\mbox{Tr}\phi^3 - \frac{3}{2}\mbox{Tr}\phi^2\Lambda + 3\Lambda^3} 
\label{i2} \end{equation}

\begin{equation} I_3 \equiv \det{\Psi} = \frac{3}
{\mbox{Tr}\phi^3 - \frac{3}{2}\mbox{Tr}\phi^2\Lambda + 3\Lambda^3} \ . \end{equation}

\noindent Since $\Lambda = \Lambda (\mbox{Tr}\phi^2, \mbox{Tr}\phi^3)$, these are 
three functions of two variables. Hence there will exist a constraint of the form

\begin{equation} h(I_1, I_2, I_3) = 0 \ . \label{constraint} \end{equation}

\noindent The form of the function $h$ will depend on the form of the function 
$\Lambda$. For the moment we pass over some technical 
difficulties that arise at this point.  

Coming back to the action (\ref{lee}), we perform a 3+1 split through $\alpha = 
(t, a)$. We define the ``magnetic'' field 

\begin{equation} B^a_i \equiv \frac{1}{2}\epsilon^{tabc}F_{bci} \equiv 
\frac{1}{2}\epsilon^{abc}F_{bci} \ . \end{equation}

\noindent We denote time derivatives with overdots. Then the canonical 
momentum is  

\begin{equation} E^a_i \equiv \frac{\partial {\cal L}}{\partial \dot{A}_{ai}} 
= \Psi_{ij}B^a_j \ . \end{equation}

\noindent One finds that the canonical Hamiltonian vanishes. Variation 
with respect to the traceless matrix $\phi$, in terms of which the matrix $\Psi$ 
is defined, gives a consistency condition which is automatically obeyed, given 
the constraints. Primary constraints arise because the matrix $\Psi$ is 
symmetric and constrained by eq. (\ref{constraint}). The first primary 
constraint is

\begin{equation} \Psi_{ij} = \Psi_{ji} \hspace{5mm} \Leftrightarrow \hspace{5mm} 
{\cal H}_a \equiv \epsilon_{abc}E^b_jB^c_k \approx 0 \ . \end{equation}

\noindent To formulate the second primary constraint, let us introduce some 
further notation. Define 

\begin{equation} EEB \equiv \epsilon_{abc}f_{ijk}E^a_iE^b_jB^c_k = 
((\mbox{Tr}\Psi )^2 - \mbox{Tr}\Psi^2 )\det{B}  
\equiv \frac{1}{6}I_1\ BBB \ . \end{equation}

\noindent In a similar notation 

\begin{equation} EBB = \frac{1}{3}I_2 \ BBB \hspace{8mm} EEE = I_3 \ BBB \ . 
\end{equation}

\noindent Then one finds 

\begin{equation} h(I_1,I_2,I_3) = 0 \hspace{5mm} \Leftrightarrow \hspace{5mm} 
{\cal H}(EEE, EEB, EBB, BBB) \approx 0 \ , \end{equation}

\noindent where ${\cal H}$ is a homogeneous function of order 1. 
Finally a secondary constraint will arise when we vary the action with respect 
to $A_{ti}$. The 3+1 decomposition is now complete, and we have arrived at the 
phase space action

\begin{equation} S = \int d^4x \ \dot{A}_{ti}E^a_i - N{\cal H} - N^a{\cal H}_a 
+ A_{ti}D_aE^a_i \ . \end{equation}

\noindent Lagrange multipliers were introduced for the primary constraints. We 
have chosen a density weight equal to $- 1$ for the lapse function $N$. 

Now vary the phase space action with respect to $E^a_i$. In most cases one can 
solve the resulting equation for the same variable, and insert the solution back 
into the phase space action. With the definitions 

\begin{equation} \eta \equiv \frac{3i}{8N\ BBB}  \end{equation}

\begin{equation} \Omega_{ij} \equiv \epsilon^{\alpha \beta \gamma \delta}
F_{\alpha \beta i}F_{\gamma \delta j} \end{equation}

\noindent (yet another three by three matrix!), one arrives in this way at 
the CDJ type \cite{Dell,Peter} action 

\begin{equation} S = \frac{1}{8} \int d^4x \ L(\mbox{Tr}\Omega, \mbox{Tr}
\Omega^2, \mbox{Tr}\Omega^3, \eta^{-1}) \ , \end{equation}

\noindent where again the function $L$ is homogeneous of order 1. In the 
calculation one must solve for the invariants of $\Psi$ as 
functions of the invariants of $\Omega$; in practice this may be difficult. 
If we impose the constraint on $\Psi$ through a Lagrange multiplier $\eta$, 
the CDJ action can be arrived at directly from the action (\ref{lee}). 

This ends our overview of the equivalence between the Pleba\'nski, CDJ, and 
Hamiltonian formulations of the models considered by Krasnov. We have hidden 
some difficulties which will appear when we consider a few examples. Let us 
start from the action (\ref{krasnov}). In eq. (\ref{kirill}) we set

\begin{equation} \Lambda = i\lambda /3 \ , \end{equation}

\noindent where $\lambda$ is a constant. It follows that 
$I_1 = 2i\lambda I_3$, and the Hamiltonian constraint is 

\begin{equation} {\cal H} = \frac{i}{2}EEB + \frac{\lambda}{3}EEE \ . 
\end{equation}
 
\noindent This is Ashtekar's Hamiltonian for GR including a cosmological 
constant \cite{Ashtekar} (and we included a constant factor for agreement 
with standard conventions). 
This establishes the equivalence between the original Pleba\'nski 
action and the Einstein-Hilbert action. The corresponding CDJ action is 
quite elegant when $\lambda = 0$ (see below), but otherwise 
one encounters a difficulty. To solve for the invariants of $\Psi$ in terms 
of the invariants of $\Omega$ one must take a square root. Consequently the 
CDJ action will have two branches---not to mention the fact that it becomes 
very complicated \cite{Peldan}.

In our second example we let the matrix $\Psi$ have constant trace. Starting 
from the action (\ref{lee}) it is easily deduced that 

\begin{equation} \mbox{Tr}\Psi = - i\alpha \hspace{5mm} \Rightarrow 
\hspace{5mm} {\cal H} = \frac{1}{2}EBB + \frac{i\alpha}{6}BBB \ . 
\end{equation}

\noindent This theory is ``dual'' to Ashtekar's. The parameter 
$\alpha$ can be set to zero through the canonical transformation 

\begin{equation} A_{ai} \rightarrow A_{ai} \hspace{10mm} E^a_i \rightarrow 
E^a_i - \frac{i\alpha}{3} B^a_i \ . \end{equation}

\noindent Let us do so for simplicity. It is clear from eq. (\ref{i2})
that there are two possibilities for deriving this model from Krasnov's action, 
namely 

\begin{equation} \Lambda = \pm \sqrt{\mbox{Tr}\phi^2/6} \ . \end{equation}

\noindent The CDJ action does not exist in this case, since the equation obtained 
by varying the phase space action with respect to $E^a_i$ is independent of 
$E^a_i$.

Our third and final example is a one-parameter family of models that 
attracted some attention because of its simple CDJ Lagrangian \cite{Capovilla,PPIB}:

\begin{equation} S = \frac{1}{8}\int d^4x \ \eta \left( \mbox{Tr}\Omega^2 + a
(\mbox{Tr}\Omega )^2\right) \ . \end{equation}

\noindent The particular case $a = - 1/2$ gives standard GR with vanishing 
cosmological constant. The Hamiltonian constraint becomes 

\begin{equation} {\cal H} = \frac{i}{2}EEB - \frac{3i(1+2a)}{4(1+ 3a)}
\frac{(EBB)^2}{BBB} \ . \end{equation}

\noindent Hence 

\begin{equation} {\cal H} = 0 \hspace{5mm} \Leftrightarrow \hspace{5mm} 
I_1 - \frac{1+2a}{1+3a}I_2 = 0 \end{equation} 

\noindent Using eqs. (\ref{i1}-\ref{i2}) we find a quadratic equation for 
the function $\Lambda$, namely 

\begin{equation} 6\Lambda = \frac{1+2a}{1+3a}\left( 3\Lambda^2 - \frac{1}{2}
\mbox{Tr}\phi^2\right) \ . \end{equation}

\noindent Consequently Krasnov's action will again have two branches. 

We have now explained the practical difficulties involved in carrying out the 
steps relating the Krasnov, CDJ, and phase space actions. But 
where is the spacetime metric in all this? Krasnov's answer is 
that the algebraic constraint on $\Sigma_{\alpha \beta i}$, obtained by varying 
the action (\ref{kirill}) with respect to $\phi$, will enable one to 
introduce a tetrad and a metric also in the case when $\Phi(\phi)$ is not traceless. 
However, the resulting metric is defined only up to an arbitrary conformal 
factor, hence the name ``non-metric gravity''.     

The Hamiltonian formulation opens another avenue to the same end. The constraint 
algebra will include the bracket  

\begin{equation} \{ {\cal H}[N], {\cal H}[M]\} = {\cal H}_a[(N\partial_aM - 
M\partial_aN)q^{ab}] \ , \label{algebra} \end{equation}

\noindent where ${\cal H}[N]$ means that the constraint has been smeared 
with a test function of the appropriate density weight. The tensor density 
$q^{ab}$ will be a definite function of the canonical variables $E^a_i$ and 
$B^a_i$. In order to agree with the usual geometric interpretation of the 
constraint algebra \cite{Hojman}, we must set   

\begin{equation} gg^{ab} \equiv q^{ab} \ , \end{equation}

\noindent where $g_{ab}$ is the spatial metric. If one then runs the ADM 
decomposition of the spacetime metric backwards, one finds (after a 
non-trivial calculation \cite{Bengtsson}) that the latter must be

\begin{equation} g_{\alpha \beta} = \frac{4}{3}\eta f_{ijk}\epsilon^
{\mu \nu \rho \sigma}F_{\alpha \mu i}F_{\nu \rho j}F_{\sigma \beta k} 
\ . \label{urbantke} \end{equation}

\noindent This is an interesting expression, since it actually implies 
that the 2-forms $F_{\alpha \beta i}$ are self dual with respect to this 
metric---and conversely \cite{Schoenberg,Urbantke}. Since Krasnov comes 
to the same conclusion, the 
two ways of defining the conformal structure agree. It should be noted 
that the conformal factor adopted in eq. (\ref{urbantke}) is conventional 
also from the Hamiltonian perspective, because one can redefine the lapse 
function with scalar functions of the canonical variables without changing 
the dynamical content of the theory---this would change the tensor density 
$q^{ab}$ in eq. (\ref{algebra}) with a factor, and hence the conformal 
factor of the metric.

If these models are to be taken seriously as alternative gravity theories, it 
must be properly explained how to make the restriction to real Lorentzian 
spacetimes, and it must be shown how to couple them to matter degrees of freedom. 
Krasnov's formulation is likely to be helpful here. If it can breathe new life 
into the asymptotic safety programme \cite{Weinberg}, it is so much the better.

\hspace{12mm}

{\bf Acknowledgement}

\hspace{5mm}

\noindent I thank Kirill Krasnov for several interesting emails.

\end{document}